\documentclass[a4paper]{jpconf}
\usepackage{graphicx}
\begin{document}
\title{NEMO-3 and SuperNEMO double beta decay experiments}

\author{A S Barabash on behalf of the NEMO Collaboration}

\address{Institute of Theoretical and Experimental Physics, \\ B. Cheremushkinskaya 25, 117218 Moscow, Russia}

\ead{Alexander.Barabash@itep.ru}

\begin{abstract}
   Latest results on $\beta\beta 0 \nu$, $\beta\beta 0 \nu\chi^0$ and $\beta\beta 2 \nu$ 
   decays 
of different isotopes from NEMO-3 double beta decay experiment are presented. In particular, 
new limits on neurtinoless
double beta decay of $^{100}$Mo and $^{82}$Se have been obtained, $T_{1/2} > 4.6\times 10^{23}$ y and  
$T_{1/2} > 1\times 10^{23}$ y (90\% C.L.), respectively. 
A possible next step with SuperNEMO detector is discussed. 
\end{abstract}

\section{Introduction}
NEMO-3 detector is currently operating in the Frejus Underground Laboratory (4800 m w.e.).
We present here recent results on $\beta\beta 0 \nu$, $\beta\beta 0 \nu\chi^0$ and $\beta\beta 2 \nu$ transitions.
At the same time the Collaboration has started to develop a SuperNEMO detector for 100 kg of $^{82}$Se.

\section{The NEMO-3 experiment}

\subsection{The NEMO-3 detector}

The experiment is based on the direct detection of two electrons produced by a double beta decay. 
NEMO-3 is able to operate with several double beta decay isotopes in the form of foils
$\sim$ 50 $\mu$m thick. The NEMO-3 detector accommodates $\sim$ 10 kg of isotopes: 6.9 kg
of $^{100}$Mo, 0.93 kg of $^{82}$Se, 0.4 kg of $^{116}$Cd, 0.45 kg of $^{130}$Te, 37 g of $^{150}$Nd, 9 g
of $^{96}$Zr and 7 g of $^{48}$Ca.
Particle detection in NEMO-3 consists of two parts. A tracking volume allows reconstruction of the
tracks of charged particles inside the detector and a calorimeter measures the energy of
$e^-$, $e^+$ and  $\gamma$ particles. The tracking part is composed of 6180 drift cells, operating in Geiger
mode, which provide three dimensional tracks. The tracking volume is surrounded by the calorimeter which
is made of 1940 blocks of plastic scintillators. In 
addition, a magnetic field of 25 Gauss parallel to the detector's axis is created by a solenoid wound
around the detector. The detector is surrounded by a passive shield.

The main characteristics of the detector's performance are the following. The energy 
resolution of the scintillation counters lies in the interval of 14-17\% (FWHM for 1 MeV 
electrons). The time resolution is 250 ps for an electron energy of 1 MeV. The reconstruction 
accuracy of a two electron (2e)  vertex is around 1 cm. The characteristics of the detector are 
studied in special calibration runs with radioactive sources. A detailed description of the detector
 and its characteristics is presented 
in \cite{ARN05}.

\subsection{Results}
\underline{Measurement of the two neutrino double beta decay.}
For $^{100}$Mo and $^{82}$Se 389 effective days of data were analyzed. In case of $^{116}$Cd, $^{96}$Zr and $^{150}$Nd 
168.4 days data were used. The results of the measurement are presented in Table 1.

\begin{center}
\begin{table}[h]
\caption{Main results on $\beta\beta 2\nu$ decay. S/B is the signal-to-background ratio. }
\centering
\begin{tabular}{@{}llll}
\br
Nuclei & Number of events & S/B ratio & $T_{1/2}$, y\\
\mr
$^{100}$Mo & 219000 & 40 & $[ 7.11 \pm 0.02(stat) \pm 0.54(syst) ] \times 10^{18}$\\
$^{82}$Se&2750&4&$[ 9.6 \pm 0.3(stat) \pm 1.0(syst) ] \times 10^{19}$\\
$^{116}$Cd&1371&7.5&$[ 2.8 \pm 0.1(stat) \pm 0.3(syst) ] \times 10^{19}$\\
$^{96}$Zr&72&0.9&$[ 2.0 \pm 0.3(stat) \pm 0.2(syst) ] \times 10^{19}$\\
$^{150}$Nd&449&2.8&$[ 9.7 \pm 0.7(stat) \pm 1.0(syst) ] \times 10^{18}$\\
\br
\end{tabular}
\end{table}
\end{center}

\begin{figure*}
\includegraphics[scale=0.3]{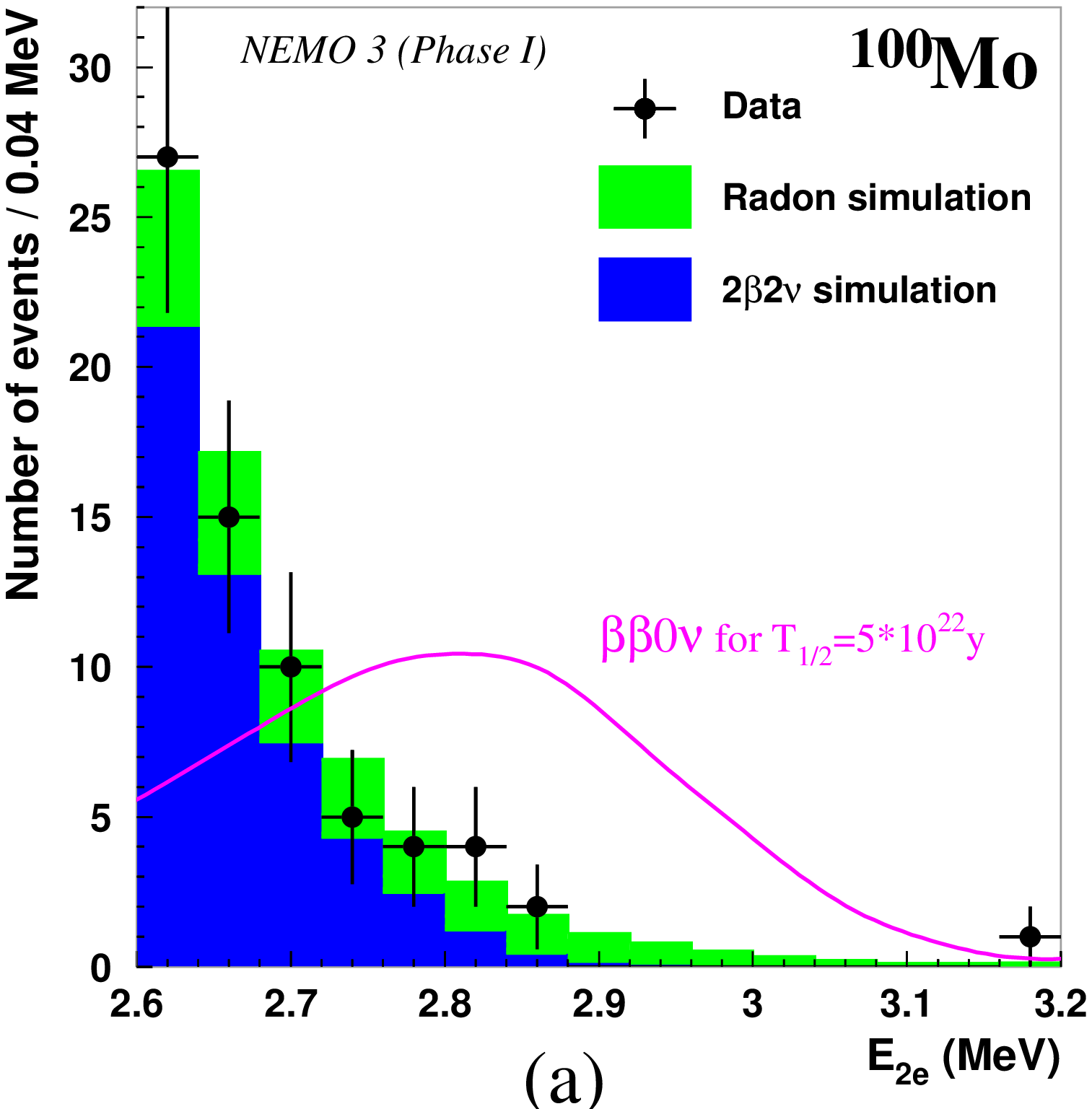}
\includegraphics[scale=0.3]{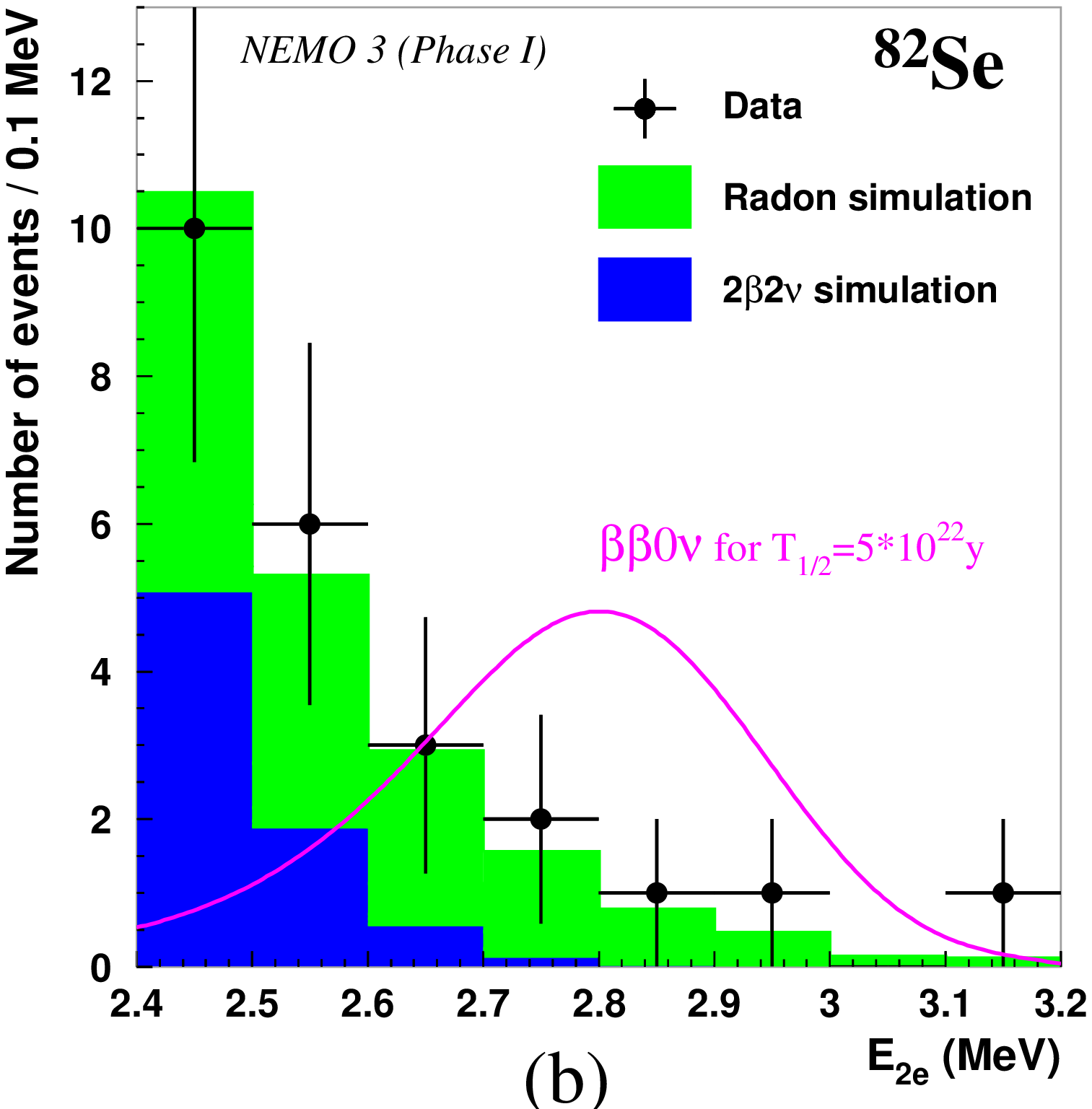}
\includegraphics[scale=0.3]{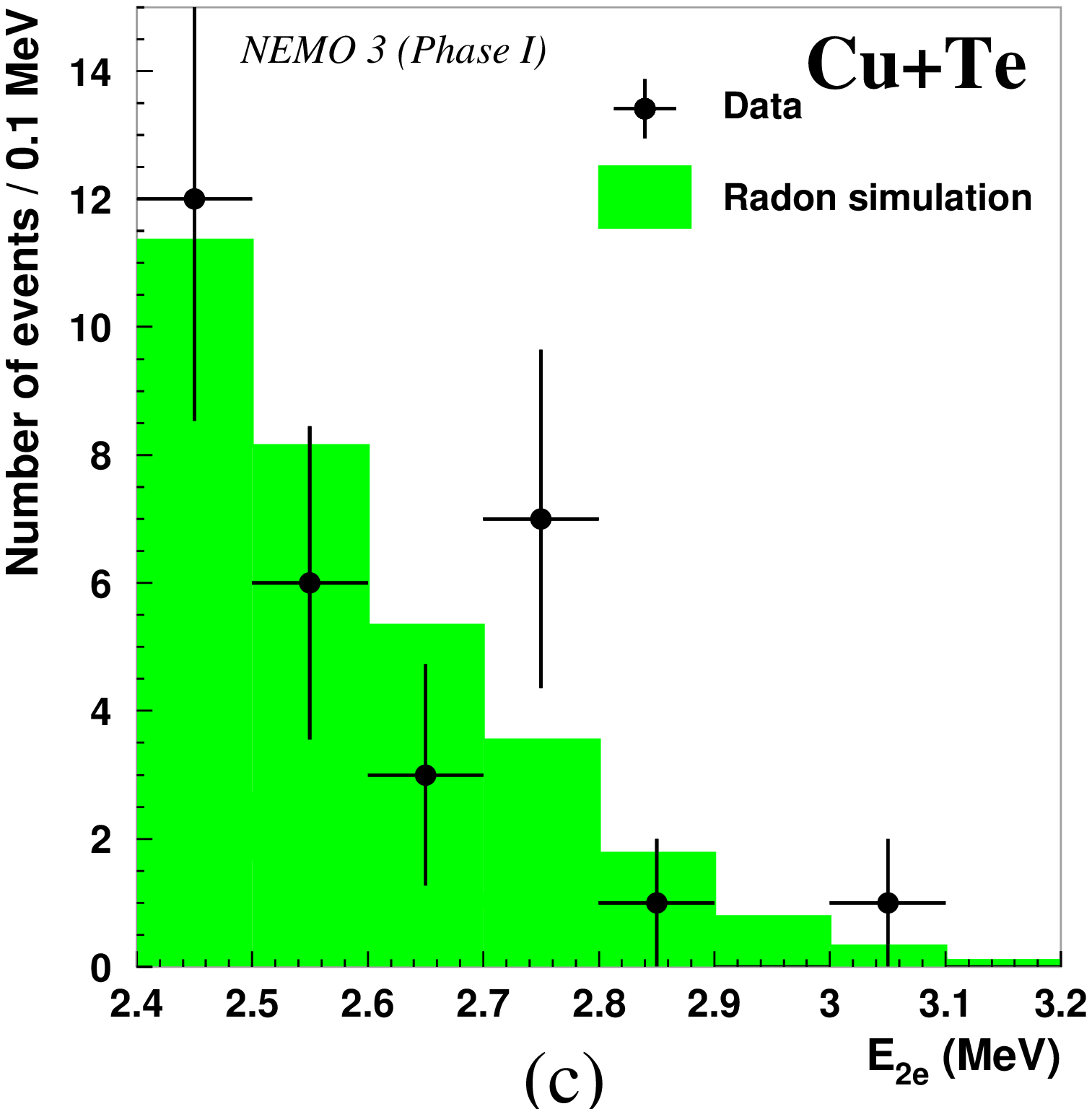}
\caption{\label{fig:results} Spectra of the energy sum of the two electrons in the $\beta\beta 0 \nu$ energy window 
after 389 effective days of data collection from February 2003 until September 2004 (Phase~I): 
(a) with  $6.914$~kg of $^{100}$Mo; (b) with $0.932$~kg of $^{82}$Se; (c) with Copper and Tellurium foils. 
The shaded histograms are the expected backgrounds computed by Monte-Carlo simulations: 
dark (blue) is the $\beta\beta2\nu$ contribution and light (green) is the Radon contribution. 
The solid line corresponds to the expected $\beta\beta 0\nu$ signal if $T_{1/2}(\beta\beta0\nu)= 5 \times 10^{22}$ y.}
\end{figure*}

\underline{Search for neutrinoless double beta decay.}
     Fig.~\ref{fig:results}(a) and (b) show the tail of the two-electron energy sum spectrum in the 
$\beta\beta 0 \nu$ energy window for $^{100}$Mo and for $^{82}$Se respectively. 
The number of 2$e^-$  events observed in the data is in agreement with the expected number of 
events from $\beta\beta 2 \nu$ decay and simulations of the radon background. 

In order to make the optimum use of all the information from the NEMO~3 detector, 
a maximum likelihood analysis has been applied to the 2$e^-$ event 
sample above 2 MeV using the three available variables: the energy sum ($E_{tot}$) of the two electrons, 
the  energy of each electron ($E_{min}$ is the minimum electron energy), and the angle between the two tracks ($cos\theta$). 
With 389 effective days of data collection, limits at 90\% C.L. obtained with the likelihood 
analysis are $T_{1/2}(\beta\beta0\nu)> 4.6 \times 10^{23}$~years  for $^{100}$Mo and 
$1.0 \times 10^{23}$~years for  $^{82}$Se. 
The corresponding upper limits for the effective Majorana neutrino mass range 
from 0.7 to 2.8 eV for $^{100}$Mo and 1.7 to 4.9 eV for $^{82}$Se depending on the nuclear matrix 
element calculation \cite{ROD05,SIM99,CIV03,STO01,CAU96}. 
For the hypothesis of a right-handed weak current, the limits at 90\% C.L. are 
$T_{1/2}(\beta\beta0\nu)> 1.7 \times 10^{23}$~years for $^{100}$Mo and 
$0.7 \times 10^{23}$~years for  $^{82}$Se, corresponding to an upper limit on the coupling 
constant of $\lambda  < 2.5 \times 10^{-6}$ for $^{100}$Mo and 
$3.8 \times 10^{-6}$ for $^{82}$Se using the nuclear calculations from references~\cite{AUN98,SUH02}.\\

\underline{Decay with Majoron emission.}
In this case, the analysis of 8023 hours of NEMO-3 data is presented. 
The limits were obtained by analyzing the deviation in the shape of the energy distribution
of the experimental data in comparison with calculated spectrum for $\beta\beta2\nu$ decay. A maximum likelihood
analysis was applied and different Majoron modes were investigated. 
The half-life limits for $^{100}$Mo and $^{82}$Se for the different decay modes are presented in Table 2.
New limits on coupling constant of Majoron to neutrino were obtained. 
In particular, 
new limits on "ordinary" Majoron 
(spectral index 1) decay of $^{100}\rm Mo$ 
and $^{82}\rm Se$ correspond to bounds  
of $\langle g_{ee} \rangle < (0.4-1.8) \cdot 10^{-4}$ 
and  $< (0.66-1.9) \cdot 10^{-4}$ using nuclear matrix element calculations from \cite{ROD05,SIM99,CIV03,STO01}.

\begin{center}
\begin{table}[h]
\caption{Limits at  90\% C.L. on $T_{1/2}$(y) for different modes of double beta decay 
with Majoron emission. n is "spectral index", which define the summed energy spectrum of the emitted electrons.}
\centering
\begin{tabular}{@{}lllll}
\br
 Nuclei & $n=1$ & $n=2$ & $n=3$ & $n=7$\\
\mr
$^{100}$Mo&$> 2.7\times10^{22}$&$> 1.7\times10^{22}$&$> 1.0\times10^{22}$&$> 7\times10^{19}$\\
$^{82}$Se&$> 1.5\times10^{22}$&$> 6.0\times10^{21}$&$> 3.1\times10^{21}$&$> 5.0\times10^{20}$\\
\br
\end{tabular}
\end{table}
\end{center}

\section{SuperNEMO experiment}
The NEMO Collaboration is currently planning a future new, bigger detector. The main idea is to use the same 
experimental technique as in NEMO-2 \cite{ARN95} and NEMO-3 \cite{ARN05} experiments and to study 100 kg
 of $^{82}$Se. The planar geometry 
and modular scheme is proposed. The energy resolution will be improved up to $\sim 8-10\%$ (FWHM) at 1 MeV and 
the efficiency
for $0\nu$ decay will be increased up to $\sim 20-40\%$. Other parameters will be the same as in case of NEMO-3.
Sensitivity of the new experiment is estimated as $\sim (1.5-2)\times10^{26}$ y for half-life or $\sim$ (0.04-0.1) eV
 for effective Majorana
neutrino mass. 
A more detailed description of the SuperNEMO detector and its 
characteristics is presented in \cite{BAR02,BAR04}.

\ack 
Portions of this work were supported 
by a grant from INTAS N 03051-3431 and a grant NATO PST.CLG.980022.


\section*{References}
 

\begin{thebibliography}{12}

\bibitem{ARN05} 
Arnold R et al. 2005 {\it Nucl. Inst. Meth.} A {\bf 536} 79
\bibitem{ROD05} 
Rodin V A et al. 2003 {\it Phys. Rev.} C {\bf 68} 044302; nucl-th/0503063
\bibitem{SIM99} 
Simkovic F et al. 1999 {\it Phys. Rev.} C {\bf 60} 055502
\bibitem{CIV03} 
Civitarese O and Suhonen J 2003 {\it Nucl. Phys.} A {\bf 729} 867
\bibitem{STO01} 
Stoica S and Klapdor-Kleingrothaus H V 2001 {\it Nucl. Phys.} A {\bf 694} 269
\bibitem{CAU96}
Caurier E et al. 1996 {\it Phys. Rev. Lett.} {\bf 77} 1954 
\bibitem{AUN98} 
Aunola M and Suhonen J 1998 {\it Nucl. Phys.} A {\bf 643}  207
\bibitem{SUH02} 
Suhonen J 2002 {\it Nucl. Phys.} A {\bf 700} 649
\bibitem{ARN95}
Arnold R et al. 1995 {\it Nucl. Inst.Meth.} A {\bf 354} 338  
\bibitem{BAR02}
Barabash A S  2002 {\it Czech. J. Phys.} {\bf 52} 575
\bibitem{BAR04} 
Barabash A S 2004 {\it Phys. At. Nucl.} {\bf 67} 1984

\end{thebibliography}
\end{document}